\documentclass[prd,superscriptaddress,amsfonts,amssymb,amsmath,showpacs,onecolumn]{revtex4-2}
\usepackage{bm}
\usepackage{amsfonts}
\usepackage{latexsym}
\usepackage{graphicx}
\usepackage{amsmath}
\usepackage{palatino}
\usepackage{mathpazo}
\usepackage{textcomp}
\usepackage{float}
\usepackage{booktabs}
\usepackage{dcolumn}
\usepackage{booktabs}
\usepackage{multirow}
\usepackage{hyperref}
\hypersetup{colorlinks,citecolor=blue}
\usepackage{amsmath}
\usepackage{xcolor}
\usepackage{orcidlink}
\usepackage{epsfig}
\usepackage{caption}
\usepackage{subcaption}
\usepackage{commath}
\usepackage{tabularx}
\def\jnl@style{\it}
\def\aaref@jnl#1{{\jnl@style#1}}

\def\aaref@jnl#1{{\jnl@style#1}}

\def\aj{\aaref@jnl{AJ}}                   
\def\apj{\aaref@jnl{ApJ}}                 
\def\apjl{\aaref@jnl{ApJ}}                
\def\apjs{\aaref@jnl{ApJS}}               
\def\apss{\aaref@jnl{Ap\&SS}}             
\def\aap{\aaref@jnl{A\&A}}                
\def\aapr{\aaref@jnl{A\&A~Rev.}}          
\def\aaps{\aaref@jnl{A\&AS}}              
\def\mnras{\aaref@jnl{Mon.~Not.~Roy.~Astron.~Soc.}}             
\def\prd{\aaref@jnl{Phys.~Rev.~D}}        
\def\prc{\aaref@jnl{Phys.~Rev.~C}}  
\def\prl{\aaref@jnl{Phys.~Rev.~Lett.}}    
\def\qjras{\aaref@jnl{QJRAS}}             
\def\skytel{\aaref@jnl{S\&T}}             
\def\ssr{\aaref@jnl{Space~Sci.~Rev.}}     
\def\zap{\aaref@jnl{ZAp}}                 
\def\nat{\aaref@jnl{Nature}}              
\def\aplett{\aaref@jnl{Astrophys.~Lett.}} 
\def\apspr{\aaref@jnl{Astrophys.~Space~Phys.~Res.}} 
\def\physrep{\aaref@jnl{Phys.~Rep.}}      
\def\physscr{\aaref@jnl{Phys.~Scr}}       
\def\commat{\aaref@jnl{Comm.~Math.~Phys.}}              
\def\science{\aaref@jnl{Science}}               
\def\cqg{\aaref@jnl{Classical Quant.~Grav.}}            
\def\jpcs{\aaref@jnl{JPCS}}                                     
\def\ijmpd{\aaref@jnl{Int.~J.~Mod.~Phys.~D}}                    
\def\grg{\aaref@jnl{Gen.~Relat.~Gravit.}}               
\def\rpp{\aaref@jnl{Rep.~Prog.~Phys.}}          
\def\npa{\aaref@jnl{Nucl.~Phys.~A}}        
\def\lrr{\aaref@jnl{Living Rev.~Rel.}}                   
\def\jcap{\aaref@jnl{J.~Cosmology Astropart.~Phys.}}    
\def\rmp{\aaref@jnl{Rev.~Mod.~Phys.}}   
\def\epjc{\aaref@jnl{Eur.~Phys.~J.~C}}

\allowdisplaybreaks[1]

\begin{document}

\color{black}       

\title{Bouncing behavior in $f(R,L_m)$ gravity: Phantom crossing and energy conditions}

\author{M. Koussour\orcidlink{0000-0002-4188-0572}}
\email[Email: ]{pr.mouhssine@gmail.com}
\affiliation{Department of Physics, University of Hassan II Casablanca, Morocco.}

\author{N. Myrzakulov\orcidlink{0000-0001-8691-9939}}
\email[Email: ]{nmyrzakulov@gmail.com}
\affiliation{L. N. Gumilyov Eurasian National University, Astana 010008,
Kazakhstan.}

\author{Javlon Rayimbaev\orcidlink{0000-0001-9293-1838}}
\email[Email: ]{javlon@astrin.uz}
\affiliation{New Uzbekistan University, Mustaqillik Ave. 54, Tashkent 100007, Uzbekistan.}
\affiliation{University of Tashkent for Applied Sciences, Gavhar Str. 1, Tashkent 100149, Uzbekistan.}
\affiliation{National University of Uzbekistan, Tashkent 100174, Uzbekistan.}

\author{Alnadhief H. A. Alfedeel\orcidlink{0000-0002-8036-268X}}%
\email[Email:]{aaalnadhief@imamu.edu.sa}
\affiliation{Department of Mathematics and Statistics, Imam Mohammad Ibn Saud Islamic University (IMSIU),\\
Riyadh 13318, Saudi Arabia.}
\affiliation{Department of Physics, Faculty of Science, University of Khartoum, P.O. Box 321, Khartoum 11115, Sudan}
\affiliation{Centre for Space Research, North-West University, Potchefstroom 2520, South Africa}

\author{H. M. Elkhair\orcidlink{0000-0000-0000-0000}}%
\email[Email:]{HMdirar@imamu.edu.sa}
\affiliation{Deanship of Scientific Research, Imam Mohammad Ibn Saud Islamic University (IMSIU), \\ P. O Bx 5701, 11432, Riyadh, Saudi Arabia}

\date{\today}
\begin{abstract}

In this work, we investigate the bouncing behavior of the universe within the framework of $f(R,L_m)$ gravity, using a simple form of $f(R,L_m)=\frac{R}{2}+L_m^\gamma$ (where $\gamma$ is a free model parameter) as previously studied. The model predicts a vanishing Hubble parameter in the early and late times, with the deceleration parameter approaching a specific limit at the bouncing point. The EoS parameter is observed to cross the phantom divide line ($\omega=-1$) near the bouncing point, indicating a significant transition from a contracting to an expanding phase. The model satisfies the necessary energy conditions for a successful bouncing scenario, with violations indicating exotic matter near the bouncing point. Stability conditions are satisfied for certain values of $\gamma$ near the bouncing point, but potential instabilities in late-time evolution require further investigation. Finally, we conclude that the $f(R,L_m)$ gravity model is promising for understanding the universe's dynamics, especially during events like the bouncing phase.


\end{abstract}

\maketitle

\section{Introduction}\label{sec2}

The significance of Einstein's general theory of relativity (GR) cannot be overstated. It stands as one of the most elegant and influential theories in all of science, revolutionizing our understanding of gravity \cite{Einstein/1916}. Unlike Newton's conception of gravity as a force emanating from objects, GR reveals gravity as a curvature in the fabric of spacetime due to the presence of matter. This profound insight fundamentally reshapes our perception of the universe. GR extends the framework of special relativity, which describes the physics of objects in uniform motion \cite{Einstein/1905}. While special relativity provides a foundation for understanding motion in the absence of gravitational forces, GR goes further by showing that acceleration and gravity are indistinguishable. This equivalence principle forms the cornerstone of GR, linking the dynamics of space-time with the distribution of matter and energy. 

Einstein's GR has garnered significant support from a range of observational evidence, cementing its status as a cornerstone of modern physics. One compelling line of evidence comes from gravitational lensing, where the gravitational field of massive objects bends and distorts light from more distant objects. This effect has been observed and studied extensively, providing direct visual confirmation of the predicted curvature of space-time \cite{Bartelmann/2001}. Another key observation supporting GR is the dynamics near the center of our Milky Way galaxy. Here, the presence of a supermassive black hole, known as Sagittarius A*, has been inferred from the highly elliptical orbits and high speeds of stars in its vicinity, consistent with GR's predictions \cite{Habibi/2017}. In addition, the Event Horizon Telescope's (EHT) groundbreaking imaging of the supermassive black hole at the center of the M87 galaxy in 2019 provided visual confirmation of GR's predictions for the shadow of a black hole \cite{EHT/2019}. Gravitational redshift, where the wavelength of electromagnetic waves is stretched as they climb out of a gravitational field, has been observed in various astrophysical contexts, directly confirming a key prediction of GR. Lastly, the detection of gravitational waves by the LIGO and Virgo collaborations \cite{Abbott1,Abbott2,Abbott3,Abbott4,Abbott5}, from the mergers of binary black holes and neutron stars, provides direct evidence for the existence of these waves and validates GR's predictions about their properties. 

While GR has proven to be a robust and versatile framework with applications across physical cosmology, it faces a significant challenge in explaining the present acceleration of the cosmos. This acceleration, observed in the expansion of the universe at the present epoch, cannot be accounted for within the framework of GR without introducing additional forms of matter-energy fields \cite{Pavlovic/2017}. This enigmatic entity is referred to as dark energy (DE), and its nature remains one of the most intriguing puzzles in cosmology. In response to the need to explain DE, various theoretical proposals have emerged, each offering a potential candidate to decode this cosmic mystery. These proposals include quintessence, a dynamic and evolving scalar field that permeates space and drives the acceleration of the universe. Tachyons, hypothetical particles that travel faster than light, have also been proposed as a candidate for DE. Theories such as f-essence and k-essence introduce complex scalar fields with specific kinetic and potential energy functions to explain the accelerating expansion. Phantom energy is a speculative form of DE with negative pressure that increases over time, leading to a future cosmic "big rip". Also, the Chaplygin gas model suggests a unified description of DE and dark matter (DM) as a single fluid with exotic properties (to explore these theories further, refer to \cite{Copeland/2006}).

Despite the compelling nature of these models, none of these theoretical candidates for DE have been directly detected or created in Earth-based laboratories. This limitation has led researchers to reconsider the concept of DE and explore modifications to the geometric aspect of Einstein's field equations. By reworking the Einstein-Hilbert action, new theories of gravity have arisen that can mimic the universe's accelerated expansion in its later stages. Some of these modified theories include $f(R)$ gravity, where $R$ is the Ricci scalar \cite{Staro/2007,Capo/2008,Chiba/2007}; $f(T)$ gravity, where $T$ is the torsion scalar \cite{Paliathanasis/2016,Salako/2013,Myrzakulov/2011}; $f(Q)$ gravity, where $Q$ is the non-metricity scalar \cite{Jim/2018,Q1,Q2,Q3,Q4,Q5,Q6,Q7,Q8}; $f(R, L_m)$ gravity, where $L_m$ is the matter Lagrangian density \cite{Harko/2010}; $f(R,G)$ gravity, where $G$ is the Gauss-Bonnet invariant \cite{Laurentis/2015,Gomez/2012}; and others. $f(R, L_m)$ gravity models, including non-minimal couplings between curvature and matter, have been extensively investigated in the literature due to their effectiveness in addressing numerous cosmological and astrophysical challenges \cite{THK-2,THK-3,V.F.-2}. In this framework of modified gravity, the energy-momentum tensor shows a non-zero covariant divergence, resulting in the appearance of an extra force perpendicular to the 4-velocities of particles. As a result, the trajectory of test particles differs from the geodesic paths predicted by GR. Furthermore, the $f(R, L_m)$ gravity does not comply with the equivalence principle and is restricted by tests conducted in the solar system. \cite{FR,JP}. Recently, there has been a notable increase in interest in exploring the intriguing cosmological consequences of the $f(R, L_m)$ gravity, with a growing number of studies addressing various aspects of this model; for example, see Refs. \cite{GM,RV-1,RV-2,Jay,THK-7,THK-8,Jaybhaye/2022,Myrzakulov/2023,Myrzakulova/2024}.

According to the principles of Big Bang cosmology, our cosmos began from a singularity, is finite in both time and space, and is approximately 13.7 billion years old. While the discovery of cosmic microwave background (CMB) radiation \cite{Penzias/1965} provides evidence supporting the Big Bang model, this model is faced with several challenges, including the flatness problem, horizon problem, and singularity problem. To address these issues, Alan Guth introduced the concept of inflation, proposing that the universe experienced a rapid, exponential expansion for an extremely short period (about $10^{-30}$ seconds) immediately following the Big Bang \cite{Guth/1981,Starobinsky/1980}. The inflationary scenario can explain many features of the CMB observations due to its adjustable parameters \cite{Brandenberger/2017}. However, while inflationary models can potentially resolve many of the aforementioned problems, they do not provide a solution to the singularity problem. As an alternative to inflationary models, attention has turned to alternative scenarios for the genesis and development of the universe, such as the cyclic universe hypothesis. This hypothesis proposes that our cosmos emerged from a previous contracting phase and is set to undergo an expanding phase without encountering a singularity; in simpler terms, it undergoes a bouncing stage. Numerous researchers \cite{Brandenberger/2017,Cai1,Cai2,Cai3,Cai4,Cai5,Cai6,Cai7} have investigated various aspects of bouncing scenarios in cosmology. These investigations include models featuring a scalar field with kinetic and potential components, a contracting universe filled with radiation, bounce models that include both DM and DE, observational studies of bouncing cosmologies using Planck data, and the features of bouncing cosmology as alternative models to inflation that align with observational evidence. Bamba et al. \cite{Bamba1,Bamba2,Bamba3,Bamba4} have explored the dynamical stability of bouncing cosmologies within the frameworks of $f(T)$, $f(G)$, and $f(R)$ gravity theories. de la Cruz-Dombriz et al. \cite{Cruz/2018} have presented a bouncing cosmology model based on teleparallel gravity. Cai et al. \cite{Cai8} have investigated bouncing models within the framework of $f(T)$ gravity. In addition, several papers explored the concept of a bouncing universe within the framework of loop quantum cosmology and modified gravity \cite{Amoros/2013,Odintsov/2014,Haro/2015,Haro/2014}. The authors investigated the possibility of a non-singular bounce in the early universe, which could address issues such as the big bang singularity. Odintsov and Oikonomou \cite{Odintsov/2015} investigated a scenario where the universe undergoes a non-singular bounce in the past and evolves towards a future singularity, providing an alternative to the standard Big Bang cosmology. Recently, Tripathy et al. \cite{Tripathy/2019} have studied specific bouncing models in the $f(R, T)$ gravity, revealing that the coupling constant between matter and geometry in the altered geometrical action has a notable impact on the cosmic dynamics around the bounce. Bajardi et al. \cite{Bajardi/2020} have investigated bouncing cosmology within the framework of $f(Q)$ symmetric teleparallel gravity. In this study, our goal is to investigate cosmological bounce scenarios within the framework of $f(R,L_m)$ gravity.

The following structure is adopted in this paper: Sec. \ref{sec2} presents an overview of $f(R,L_m)$ gravity. Sec. \ref{sec3} introduces the concept of cosmological bouncing solutions, defines a parametrization of the bouncing scale factor, and examines the detailed dynamical evolution of the universe. Sec. \ref{sec4} investigates the violation of energy conditions. In Sec. \ref{sec5}, we analyze the stability of our model by using the sound velocity. Finally, Sec. \ref{sec6} presents a summary and conclusions.

\section{Field equations in $f(R, L_m)$ gravity}\label{sec2}

For the formalism of $f(R, L_m)$ gravity, the modified action is expressed as follows \cite{Harko/2010}:
\begin{equation}\label{Action}
S= \int{\sqrt{-g}d^4x f(R,L_m)}. 
\end{equation}

Here, we adopt the convention $8 \pi G=c=1$, where $G$ and $c$ represent the Newtonian gravitational constant and the speed of light, respectively. $g$ represents the metric determinant, $R$ stands for the Ricci scalar, and $L_m$ denotes the matter Lagrangian density associated with the energy-momentum tensor, which can be expressed as
\begin{equation}\label{EMT}
\mathcal{T}_{\mu\nu} = \frac{-2}{\sqrt{-g}} \frac{\delta(\sqrt{-g}L_m)}{\delta g^{\mu\nu}}.
\end{equation}

The field equations governing $f(R, L_m)$ gravity are derived by varying the action $S$ (as defined in Eq. (\ref{Action})) with respect to the metric tensor $g_{\mu\nu}$, resulting in:
\begin{equation}\label{FE}
f_R R_{\mu\nu} + (g_{\mu\nu} \square - \nabla_\mu \nabla_\nu)f_R - \frac{1}{2} (f-f_{L_m}L_m)g_{\mu\nu} = \frac{1}{2} f_{L_m} \mathcal{T}_{\mu\nu}.
\end{equation}

Here, we define $f_R \equiv \frac{\partial f}{\partial R}$ and $f_{L_m} \equiv \frac{\partial f}{\partial L_m}$ as our notations. Furthermore, after calculating the covariant derivative of Eq. \eqref{FE}, we derive
\begin{equation}\label{CD}
\nabla^\mu \mathcal{T}_{\mu\nu} = 2\nabla^\mu ln(f_{L_m}) \frac{\partial L_m}{\partial g^{\mu\nu}}.
\end{equation}

Next, we assume a homogeneous and isotropic universe described by the FLRW (Friedmann-Lema\^{i}tre-Robertson-Walker) metric, which is characterized by the line element \cite{Ryden/2003}:
\begin{equation}\label{FLRW}
ds^2= -dt^2 + a^2(t)[dx^2+dy^2+dz^2],
\end{equation}
where $a(t)$ is referred to as the cosmic scale factor. Now, we consider the matter Lagrangian $L_m=\rho$ \cite{Harko/2015} for a perfect fluid, where the energy-momentum tensor is represented by
\begin{equation}\label{EMT1}
\mathcal{T}_{\mu\nu}=(\rho+p)u_\mu u_\nu + pg_{\mu\nu},
\end{equation}
where $u^\mu=(1,0,0,0)$ represents the components of the 4-velocity vector for the perfect fluid, satisfying the following relationship: $u^{\mu}u_{\mu}=-1$. Also, $\rho$ and $p$ represent the energy density and pressure in the universe, respectively. Using the expression for the energy-momentum tensor, the trace $\mathcal{T}$ can be found as
\begin{equation}
   \mathcal{T}=g^{\mu \nu} \mathcal{T}_{\mu \nu}=3p-\rho,
\end{equation}
and the Ricci scalar takes the value of
\begin{equation}
R= 6 ( \dot{H}+2H^2 ),
\end{equation}
where $H=\frac{\dot{a}}{a}$ represents the Hubble parameter, which characterizes the rate at which the universe is expanding.

Therefore, the modified Friedmann equations, governing the universe's dynamics in $f(R, L_m)$ gravity, can be expressed as \cite{Jaybhaye/2022}
\begin{equation}\label{F1}
3H^2 f_R + \frac{1}{2} \left( f-f_R R-f_{L_m}L_m \right) + 3H \dot{f_R}= \frac{1}{2}f_{L_m} \rho,
\end{equation}
and
\begin{equation}\label{F2}
\dot{H}f_R + 3H^2 f_R - \ddot{f_R} -3H\dot{f_R} + \frac{1}{2} \left( f_{L_m}L_m - f \right) = \frac{1}{2} f_{L_m}p,
\end{equation}  
where the dots represent derivatives with respect to cosmic time $t$. 

For our investigation, we consider the following functional form in the context of $f(R, L_m)$ gravity \cite{Jaybhaye/2022,Myrzakulov1,Myrzakulova,Harko/2014}
\begin{equation}\label{fRL} 
f(R,L_m)=\frac{R}{2}+L_m^\gamma,
\end{equation}
where $\gamma$ is a free parameter. The model we are considering is quite general, inspired by the functional form $f(R, L_m)= f_{1}(R) + f_{2}(R) G(L_m)$, which signifies a general coupling between matter and geometry \cite{Harko/2014}. It is noteworthy that when $\gamma=1$, the equations revert to the standard Friedmann equations of GR. In this scenario, the modified Friedmann equations \eqref{F1} and \eqref{F2} can be written as:
\begin{eqnarray}\label{F11}
3H^2&=&(2\gamma-1) \rho^{\gamma}, \\
\label{F22}
2\dot{H}+3H^2&=&\left[(\gamma-1)\rho-\gamma p\right]\rho^{\gamma-1}.
\end{eqnarray}

Using Eqs. (\ref{F11}) and (\ref{F22}), we obtain the equation of state (EoS)
parameter $\omega=\frac{p}{\rho}$ as
\begin{equation}
    \omega=-1+\frac{(2-4 \gamma ) \dot{H}}{3 \gamma  H^2}.
    \label{EoS}
\end{equation}

The dynamic evolution of physical parameters such as the EoS parameter, depends on the evolution of the Hubble parameter and $\gamma$. The EoS parameter reduces to that of GR when $\gamma=1$. In general, to analyze the behavior of cosmological parameters, an additional constraint needs to be imposed on the Hubble parameter. In the next section, we will introduce a bounce scenario as an additional constraint.

\section{Cosmological bouncing solutions}
\label{sec3}

Recently, the concept of a bouncing scenario has gained traction as an alternative to the Big Bang theory. This scenario proposes that the universe experiences a big bounce rather than a Big Bang. In this cosmological model, the big bounce can be seen as an oscillating or cyclic occurrence, with each cosmological cycle beginning with the collapse of the preceding universe. This model suggests that the universe initially contracts to a finite volume before undergoing expansion once more. This concept provides a potential resolution to the singularity problem encountered in the standard Big Bang cosmology within the framework of GR. In this paper, we focus on the matter bounce cosmological model, which has been studied extensively in the early-stage evolution of the universe and is derived from loop quantum cosmology \cite{Amoros/2013,Odintsov/2014,Haro/2015,Haro/2014}. As the cosmic time approaches the bounce time $t_b$, the scale factor, Hubble parameter, energy density, and pressure remain finite.

For a bounce to be successful, it is noted that a violation of the null energy condition (NEC) is necessary for a period near the bounce within the framework of the FLRW metric. Furthermore, the EoS parameter $\omega$ of the universe's matter content must undergo a transition from $\omega<-1$ to $\omega>-1$ to enter the Big Bang era following the bounce \cite{Cai9,Cai10}. Observational evidence favors the quintom model, which has been suggested to examine the dynamics of DE with EoS parameter $\omega>-1$ in the past and $\omega<-1$ at current era \cite{Zhao/2007,Feng/2005}. The quintom model is a non-static model of DE that behaves differently from other DE models such as the quintessence, and phantom, influencing the determination of universe evolution. The comprehensive explanation of the essential conditions required to achieve a successful bouncing of the universe in standard cosmology is as follows \cite{Cai9}:
\begin{itemize}
    \item In a contracting universe, the scale factor $a(t)$ is decreasing, denoted as $a(t) < 0$, while in an expanding universe, the scale factor $a(t)$ is increasing, denoted as $a(t) > 0$. At the transfer point, the scale factor reaches a non-zero minimum value. This type of bouncing scenario naturally avoids the singularity that is unavoidable in the standard cosmology. In simpler terms, during the bouncing phase, $\dot{a}(t) > 0$ for some time in the vicinity of the bounce point.
    \item Secondly, the Hubble parameter $H(t)$ transitions from negative values ($H<0$) when the universe contracts to positive values ($H>0$) when the universe expands, with $H=0$ representing the bouncing point. In standard cosmology, for a bouncing model to be successful, it is necessary that $\dot H=-4\pi G\rho (1+\omega)>0$ in the vicinity of a bouncing point. This condition is equivalent to the violation of the NEC in GR. This equation shows that $\omega<-1$ near the bouncing point.
    \item Finally, the EoS parameter $\omega$ crosses the phantom divide line ($\omega=-1$), a notable feature of the quintom model.
\end{itemize}

Based on the aforementioned bouncing scenario and the model-independent approach to studying cosmological models \cite{Pacif1,Pacif2}, this paper focuses on highlighting the Hubble parameter $H(t)$ (or equivalently, the scale factor $a(t)$), which describes the universe's expansion. Our goal is to demonstrate how this parameter can lead to significant bouncing solutions to the modified Friedmann equations in $f(R,L_m)$ gravity. In this work, we investigate the parameterization of the scale factor during the bouncing phase as \cite{Navo/2020,Singh/2023}
\begin{equation}
    a(t)=(\alpha+\beta t^2)^{\frac{1}{n}},
    \label{SF}
\end{equation}
where $\alpha>0$, $\beta>0$, and $n>0$ are constants. The corresponding Hubble parameter is expressed as
\begin{equation}
    H(t)=\frac{\dot a}{a}=\frac{2 \beta  t}{n(\alpha+\beta t^2)}.
    \label{HP}
\end{equation}

Furthermore, using the relation $q=-1-\frac{\dot H}{H^2}$, the deceleration parameter can be calculated as
\begin{equation}
    q(t)=\frac{1}{2} \left(-\frac{\alpha  n}{\beta  t^2}+n-2\right).
    \label{DP}
\end{equation}

\begin{figure}[h]
     \centering
     \begin{subfigure}[b]{0.33\textwidth}
         \centering
         \includegraphics[width=\textwidth]{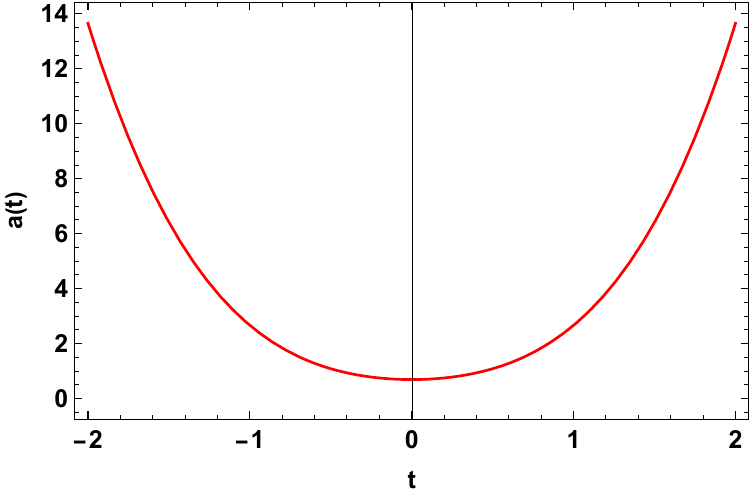}
         \caption{Scale factor}
         \label{F_a}
     \end{subfigure}
     \hfill
     \begin{subfigure}[b]{0.33\textwidth}
         \centering
         \includegraphics[width=\textwidth]{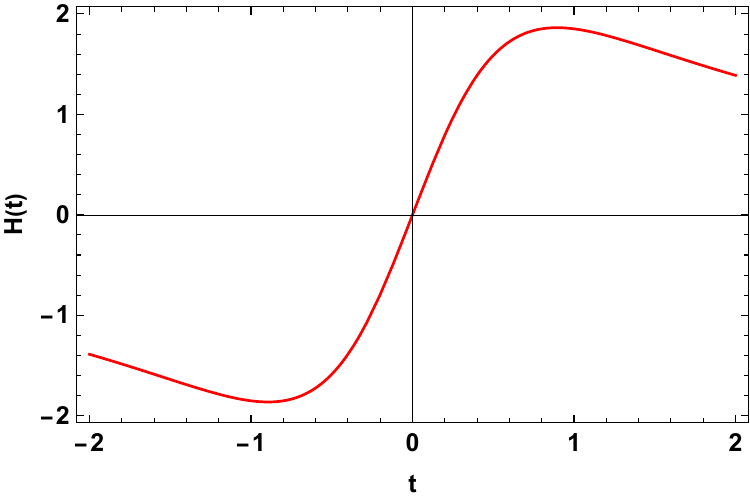}
         \caption{Hubble parameter}
         \label{F_H}
     \end{subfigure}
     \hfill
     \begin{subfigure}[b]{0.33\textwidth}
         \centering
         \includegraphics[width=\textwidth]{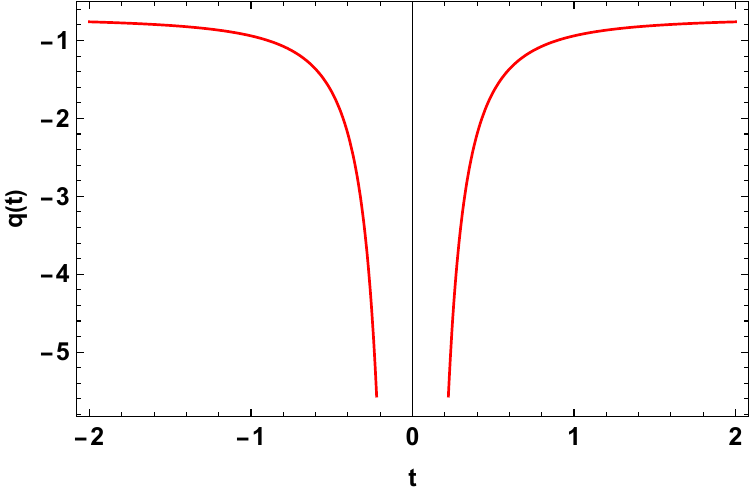}
         \caption{Deceleration parameter}
         \label{F_q}
     \end{subfigure}
        \caption{The behavior of the scale factor, Hubble parameter, and deceleration parameter as functions of cosmic time.}
        \label{F_bou}
\end{figure}

From Eqs. (\ref{HP}) and (\ref{DP}), it is found that the Hubble parameter vanishes in the early and late times, while the deceleration parameter approaches the limiting value $\frac{(n-2)}{2}$. At the bouncing point, $H(t)=0$ and $q(t)$ diverges. Since $\alpha$, $\beta$, and $n$ are positive, we fix their values at $0.8$, $1.0$, and $0.6$, respectively, to examine the cosmological parameters. In Fig. \ref{F_bou}, the plots of the scale factor, Hubble parameter, and deceleration parameter exhibit behavior consistent with the prescribed bouncing conditions. The favorable agreement between the theoretical predictions and the bouncing conditions enhances our confidence in the validity and predictive power of the model. Specially, at the bouncing point, the scale factor reaches its minimum value before starting to increase, as shown in Fig. \ref{F_a}. In Fig. \ref{F_H}, the Hubble parameter initially takes on a negative value, indicating a contracting phase ($H<0$). It then crosses the bouncing point ($H=0$) and transitions into an expanding phase ($H>0$). The behavior of the deceleration parameter is depicted in Fig. \ref{F_q}. It diverges at the singularity corresponding to the bouncing point and approaches the asymptotic value of $q=-1$ at late times. These dynamics illustrate the significant shifts in the evolution of the Universe from contraction to expansion, capturing essential features of the bouncing cosmological model under consideration.

The EoS parameter is a highly valuable metric for assessing the feasibility of a bouncing scenario. From Eqs. (\ref{EoS}) and (\ref{HP}), we have
\begin{equation}
    \omega(t)=-1+\frac{(2 \gamma-1) n \left(\beta  t^2-\alpha \right)}{3 \beta  \gamma t^2}.
\end{equation}

\begin{figure}[H]
\centering
\includegraphics[scale=0.7]{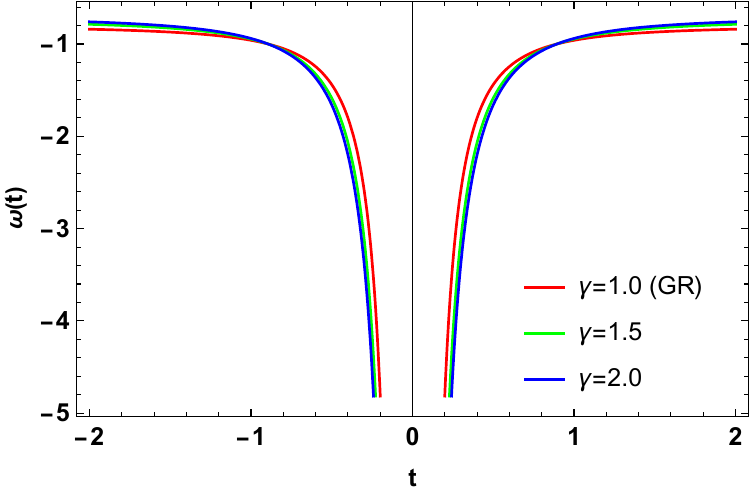}
\caption{The behavior of the EoS parameter as functions of cosmic time for three values of $\gamma$.}\label{F_EoS}
\end{figure}

As mentioned, the EoS parameter plays a crucial role in understanding the dynamics of the universe, particularly during significant cosmological events like the bouncing phase. In the context of a bouncing cosmological model, the EoS parameter governs the relationship between the energy density and pressure of the cosmic fluid. During the contracting phase leading up to the bounce, the EoS parameter can exhibit behaviors that deviate from those predicted by classical matter or radiation-dominated eras. For instance, in scenarios where the EoS parameter crosses the phantom divide line ($\omega=-1$), indicating a transition from a phase with high negative pressure to one with low negative pressure, the Universe undergoes a remarkable transition from contraction to expansion. This transition is characterized by the EoS parameter crossing $\omega=-1$ at the bouncing point, signaling a shift in the dominant energy components driving the cosmic evolution. Understanding the behavior of the EoS parameter during the bouncing phase is essential for constructing viable cosmological models that can account for the observed features of the Universe's expansion history. In this model (see Fig. \ref{F_EoS}), it is observed that the EoS parameter crosses the phantom divide line ($\omega=-1$) in the vicinity of the bouncing point at $t=0$ for all three positive values of $\gamma$, specifically $\gamma=1.0$ corresponding to the case of GR, $\gamma=1.5$, and $\gamma=2.0$. This crossing is a significant criterion for a successful bouncing model \cite{Cai10}. After crossing the phantom divide line, $\omega$ transitions from the phantom region (where $\omega<-1$) to the quintessence region (where $\omega>-1$) at late times. This transition in $\omega$ signifies a change in the nature of the cosmic fluid, from one with high negative pressure (associated with phantom energy) to one with low negative pressure (associated with quintessence), indicating a shift in the dominant energy components driving the cosmic evolution. Finally, our choice of positive values for the parameter $\gamma$ in our analysis is motivated by observational constraints and theoretical considerations \cite{Jaybhaye/2022,Myrzakulov/2023}. However, we have also explored the behavior of the EoS parameter $\omega$ when $\gamma$ is taken as negative. Surprisingly, we find that the behavior of $\omega$ remains consistent with our findings for positive $\gamma$. This suggests that our model's predictions for the crossing of the phantom divide line ($\omega=-1$) near the bouncing point are robust and not strongly dependent on the specific value of $\gamma$ within a certain range.

\section{Violation of energy conditions}
\label{sec4}

The energy conditions (ECs) are straightforward restrictions on various linear combinations of energy density and pressure. They impose restrictions such as the non-negativity of energy density and the always attractive nature of gravity \cite{Visser/2000}. These conditions emerged from the Raychaudhuri equation \cite{Carroll/2004}. The various ECs, including the null energy condition (NEC), weak energy condition (WEC), dominant energy condition (DEC), and strong energy condition (SEC), are defined as
\begin{itemize}
\item NEC: $\rho+p\geq 0\,$;

\item WEC: $\rho\geq 0$ and $\rho+p\geq 0\,$;

\item DEC: $\rho\geq 0$ and $|p|\leq \rho\,$.

\item SEC:  $\rho+3p\geq 0\,$.
\end{itemize}

\begin{figure}[H]
     \centering
     \begin{subfigure}[b]{0.33\textwidth}
         \centering
         \includegraphics[width=\textwidth]{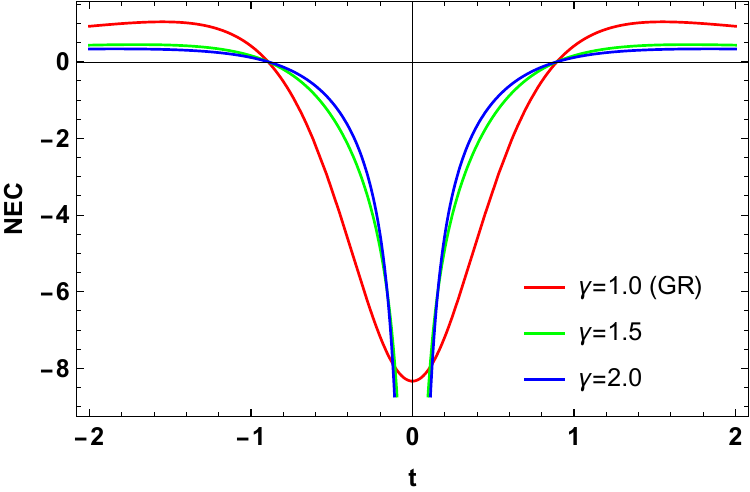}
         \caption{NEC ($\rho+p>0$)}
         \label{F_NEC}
     \end{subfigure}
     \hfill
     \begin{subfigure}[b]{0.33\textwidth}
         \centering
         \includegraphics[width=\textwidth]{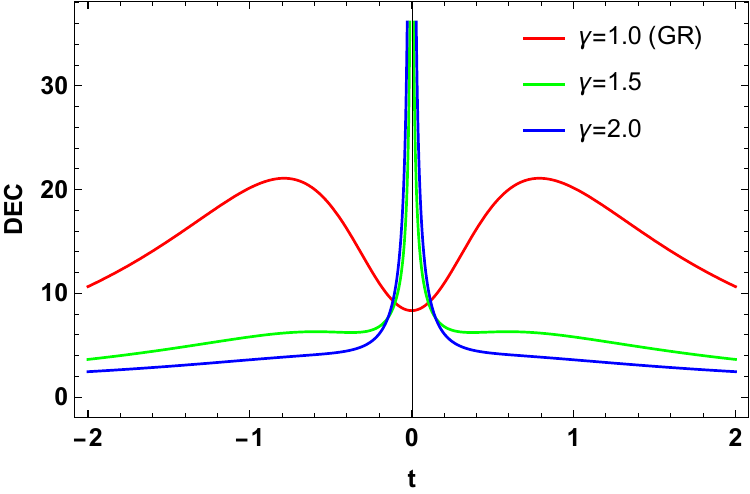}
         \caption{DEC ($\rho-p>0$)}
         \label{F_DEC}
     \end{subfigure}
     \hfill
     \begin{subfigure}[b]{0.33\textwidth}
         \centering
         \includegraphics[width=\textwidth]{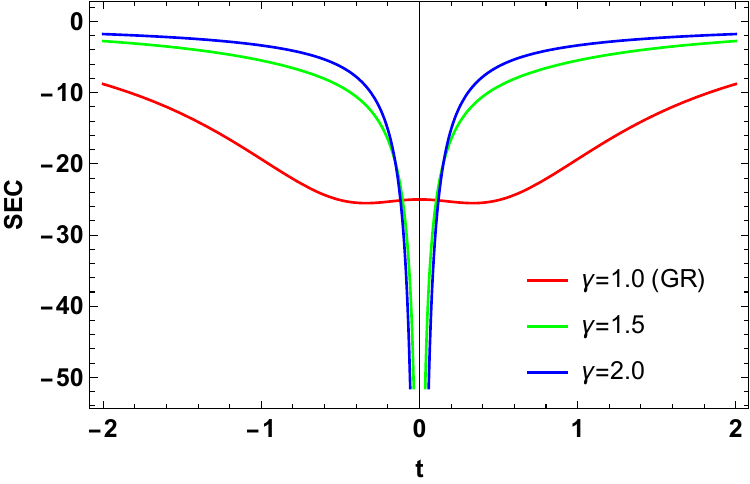}
         \caption{SEC ($\rho+3p>0$)}
         \label{F_SEC}
     \end{subfigure}
        \caption{The behavior of the ECs as functions of cosmic time for three values of $\gamma$.}
        \label{F_ECs}
\end{figure}

To achieve a successful bouncing scenario, it is necessary that $\dot H=-4\pi G\rho (1+\omega)>0$ in the vicinity of the bouncing point. Therefore, it is crucial to analyze the variations of the NEC with respect to time in the plots of Fig. \ref{F_ECs}. From Fig. \ref{F_NEC}, we observe that the NEC ($\rho+p>0$) is violated at the bouncing point, which is a favorable outcome for a successful bouncing scenario. This violation indicates that the ECs necessary for the bouncing model are satisfied, providing further support for the credibility of the proposed cosmological model. In Fig. \ref{F_SEC}, it is evident that the SEC ($\rho+3p>0$) is violated in the vicinity of the bouncing point at $t=0$, while the DEC ($\rho-p>0$) remains valid (see Fig. \ref{F_DEC}). The violation of the SEC suggests the presence of exotic matter in the universe. Additionally, the DEC is upheld in our model. Thus, our model meets all the necessary criteria for a bouncing model in $f(R,L_m)$ gravity.

\section{Stability Analysis} \label{sec5}

This section focuses on analyzing the stability of modified $f(R,L_m)$ gravity as presented. For these reasons, we consider the universe to be filled with a perfect fluid, allowing us to define the adiabatic speed of sound:
\begin{equation}
C_s^2 = \frac{dp}{d\rho}.
\end{equation}

For a system to be thermodynamically or mechanically stable, the sound velocity $C_s^2$ should remain positive, indicating stability. In addition, to ensure mechanical stability, $C_s^2$ should not exceed 1. Thus, the region where $0 \leq C_s^2 \leq 1$ corresponds to stable solutions. The violation of the NEC can lead to the emergence of ghost fields, which may result in potentially dangerous instabilities at both classical and quantum levels \cite{Dubovsky/2006}. Furthermore, complete avoidance of superluminality may not be achievable. Previous studies have shown that a regular bounce with spatial sections can occur only when the NEC $\rho + p > 0$ is violated in the context of gravitational theory \cite{Tripathy/2021}. This is the main reason for using negative energy scalar fields with ghost condensates \cite{Arkani/2004}, conformal galileon \cite{Elder/2014}, and other techniques to achieve bounces, which may occasionally result in instabilities requiring further investigation \cite{Creminelli/2006}. Recent studies have explored non-singular early universe paradigms within a generic scalar-tensor theory, such as the DHOST bounce framework, where instability can be avoided \cite{Ilyas/2020,Ilyas/2021,Zhu/2021}. 

For our bouncing cosmological model, we get
\begin{equation}
C_s^2(t) =\frac{1}{3} \left[\frac{\alpha  (\gamma-1) (2 \gamma-1) n}{\beta  \gamma t^2}+\frac{2 \alpha  (2 \gamma-1) n}{\alpha -\beta  t^2}+\left(2-\frac{1}{\gamma}\right) n-3\right].
\end{equation}

\begin{figure}[H]
\centering
\includegraphics[scale=0.7]{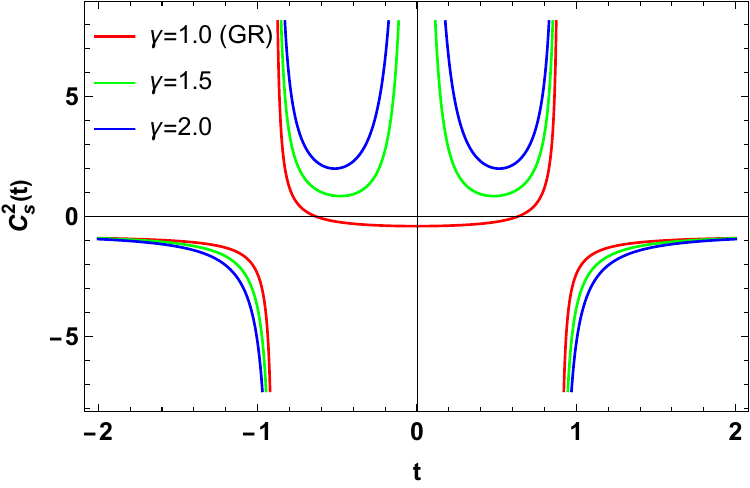}
\caption{The behavior of the sound velocity as functions of cosmic time for three values of $\gamma$.}\label{F_vs}
\end{figure}

In Fig. \ref{F_vs}, the stability of the $f(R,L_m)$ gravity model is depicted. In the vicinity of the bouncing point, the stability condition is satisfied for the values of $\gamma=1.5$ and $\gamma=2.0$, while for GR (where $\gamma=1.0$), the stability condition is not satisfied. Furthermore, in the late-time evolution, the stability condition is not met, as $C_s^2$ remains negative. These findings suggest that the model may exhibit some kind of instability, especially in the late stages of cosmic evolution, which warrants further investigation.

\section{Conclusion}\label{sec6}

In this work, we have investigated the bouncing behavior of the universe within the framework of $f(R,L_m)$ gravity. We have adopted a straightforward form of $f(R,L_m)$ as previously studied in the literature. Specially, we assume the functional form $f(R,L_m)=\frac{R}{2}+L_m^\gamma$, where $\gamma$ is a free model parameter, and then applied the model-independent approach by parameterizing the scale factor with respect to cosmic time. The analysis of the bouncing cosmological model within the framework of $f(R,L_m)$ gravity yields important insights into the dynamics and evolution of the universe. Key findings and conclusions can be summarized as follows:
\begin{itemize}
    \item \textbf{Scale factor, Hubble parameter, and deceleration parameter:} The behavior of the scale factor, Hubble parameter, and deceleration parameter exhibits characteristics consistent with the prescribed bouncing conditions. The model predicts a vanishing Hubble parameter in the early and late times, with the deceleration parameter approaching the limiting value $\frac{(n-2)}{2}$. At the bouncing point, $H(t)=0$ and $q(t)$ diverges. The chosen values of $\alpha=0.8$, $\beta=1.0$, and $n=0.6$ provide a favorable agreement between theoretical predictions and bouncing conditions, enhancing the model's validity and predictive power.
    \item \textbf{EoS parameter:} The model predicts that the EoS parameter crosses the phantom divide line ($\omega=-1$) in the vicinity of the bouncing point, signaling a transition from a phase with high negative pressure to one with low negative pressure. This transition is crucial for the shift from contraction to expansion, capturing essential features of the bouncing cosmological model. 
    \item \textbf{Violation of ECs:} The model satisfies the necessary ECs for a successful bouncing scenario. The NEC is violated at the bouncing point, indicating that the energy conditions required for the bouncing model are satisfied. The SEC is violated near the bouncing point, suggesting the presence of exotic matter in the universe. However, the DEC is upheld in the model.
    \item \textbf{Stability analysis:} The stability analysis of the $f(R,L_m )$ gravity model indicates that stability conditions are satisfied for certain values of $\gamma$ in the vicinity of the bouncing point. However, in the late-time evolution, the stability condition is not satisfied, suggesting potential instabilities that require further investigation.
\end{itemize}

In conclusion, the $f(R,L_m )$ gravity model offers a promising framework for understanding the dynamics of the Universe, particularly during cosmological events such as the bouncing phase. The model's ability to reproduce key features of cosmic evolution, such as the transition from contraction to expansion and the violation of energy conditions, highlights its potential to provide new insights into fundamental aspects of cosmology.

\section*{Acknowledgments}
The authors extend their appreciation to the Deputyship for Research \& Innovation, Ministry of Education in Saudi Arabia for funding this research through project number IFP-IMSIU-2023110. The authors also appreciate the Deanship of Scientific Research at Imam Mohammad Ibn Saud Islamic University (IMSIU) for supporting and supervising this project.

\section*{Data Availability Statement}
There are no new data associated with this article.


\end{document}